\documentclass[useAMS,usenatbib]{mn2e}
\usepackage[dvips]{graphicx,color}

\usepackage{amssymb}
\usepackage{indentfirst}
\usepackage{epsfig}
\usepackage{pdflscape}
\usepackage{afterpage}

\newcommand{\Ka}{\ensuremath{\hbox{K}\alpha~}}

\def\sw{{\it Swift}}
\def\fm{{\it Fermi}}

\begin{document}
\title[]{Super-Eddington accretion of the first Galactic Ultra-luminous X-ray pulsar
Swift J0243.6+6124}

\author[J. Liu et al.]{Jiren Liu$^{1}$\thanks{E-mail: liujiren@bjp.org.cn},
Peter A. Jenke$^2$,  Long Ji$^3$, Shuang-Nan Zhang$^4$, Shu Zhang$^4$,
\newauthor
Mingyu Ge$^4$, Jinyuan Liao$^4$, Xiaobo Li$^4$, Liming Song$^4$\\
	 $^{1}$Beijing Planetarium, 138 Sizhimenwai Road, Beijing 100044, China\\
	 $^{2}$University of Alabama in Huntsville, Huntsville, AL 35812, USA\\
	 $^{3}$School of Physics and Astronomy, Sun Yat-sen University, 2 Daxue Road, Zhuhai, Guangdong 519082, China\\
	 $^{4}$Institute of High Energy Physics, Chinese Academy of Sciences
	 Beijing 100049, China
 }

\date{}

\maketitle

\begin{abstract}
We present a detailed timing study of the pulse profile 
of Swift J0243.6+6124 with HXMT and \fm/GBM data during its 2017 giant outburst. 
The double-peak profile at luminosity above $5\times10^{38}$erg\,s$^{-1}$ 
is found to be 0.25 phase offset from that below $1.5\times10^{38}$erg\,s$^{-1}$, 
which strongly supports for a transition from a pencil beam to a fan beam, and thus 
for the formation of shock dominated accretion column.
During the rising stage of the high double-peak regime, the faint peak 
got saturated in 10-100 keV band above a luminosity of
$L_t\sim1.3\times10^{39}$erg\,s$^{-1}$, 
which is coincident with sudden spectral changes of both the main and faint peaks.
They imply a sudden change of emission pattern around $L_t$.
	The spin-up rate ($\dot{\nu}$) is linearly correlated with luminosity ($L$) below 
$L_t$, consistent with the prediction of a radiation pressure dominated (RPD) disk.
The $\dot{\nu}-L$ relation flattens above $L_t$,
indicating a less efficient transfer of angular momentum and a change of
accretion disk geometry above $L_t$.
It is likely due to irradiation of the disk by the central
accretion column and indicates significant radiation feedback
before the inner disk radius reaching the spherization radius.
\end{abstract}


\section{Introduction}

Recent discovery of ultra-luminous X-ray pulsars (ULXP) in nearby galaxies revealed
the existence of accretion onto magnetized neutron stars with observed luminosity of 5-500 times
their Eddington limit
($\sim2\times10^{38}$erg\,s$^{-1}$) \citep[e.g.][]{Bac14,Fur16,Isr17a,Isr17b,Car18}.
How the accretion proceeds in such strong radiation
fields and how the strong radiation affects the accretion process
are unclear yet \citep[e.g.][]{Isr17a,King20,Mus21}.

Swift J0243.6+6124 (J0243 hereafter) is a new Be-type X-ray binary 
discovered by \sw\  in October 2017 during one of the brightest outburst \citep{Cen17}.
A pulse period around 9.8 s was detected \citep{Ken17,Jen17} with an orbital
period around 27.7 days \citep{Ge17, Jen18, Dor18}. The optical star was identified as 
a O9.5Ve star \citep{Kou17,Bik17,Rei20}. The peak flux of Swift J0243 within \sw/BAT band 
reached $\sim9$ Crab. With a distance of 6.8 kpc based on {\it Gaia} data \citep{Bai18}, 
the peak luminosity of Swift J0243 is $\sim2\times10^{39}$erg\,s$^{-1}$, making it 
the first Galactic ULXP \citep[e.g.][]{Tsy18,Wil18}.

Significant evolution of temporal and spectral properties of Swift J0243 was observed
during the giant outburst.
Its pulse profile was double-peaked at low fluxes 
and changed to a single peak above a luminosity $\sim1.5\times10^{38}$erg\,s$^{-1}$,
and changed again to double peaks above a luminosity $\sim5\times10^{38}$erg\,s$^{-1}$, as 
revealed by \fm/GBM and {\it Insight}-HXMT data in high energy band \citep{Wil18,Dor20}.
A similar single-to-double transition around 
$5\times10^{38}$erg\,s$^{-1}$ was also found in low energy band \citep{Tsy18, Wil18, Sug20}.
The aperiodic power spectrum density (PSD) changed with fluxes \citep{Wil18,Dor20}.
The spectral properties of Swift J0243 were also found to change around 
a luminosity of $1-5\times10^{38}$erg\,s$^{-1}$
\citep[e.g.][]{Wil18,Sug20,Kong20,Wang20}. 

The transition of these temporal and spectral properties has been generally attributed 
to formation of an accretion column. 
\citet{Dor20} proposed that the transition 
at $\sim5\times10^{38}$erg\,s$^{-1}$ could be associated with the transition 
of a gas-state accretion disk to a radiation pressure dominated (RPD) state.
In this {\it Letter} we perform a phase-coherent
pulse profile time evolution study of Swift J0243 with HXMT and GBM data to explore
how an accretion column evolves with time and how it affects the accretion process.

\begin{figure*}
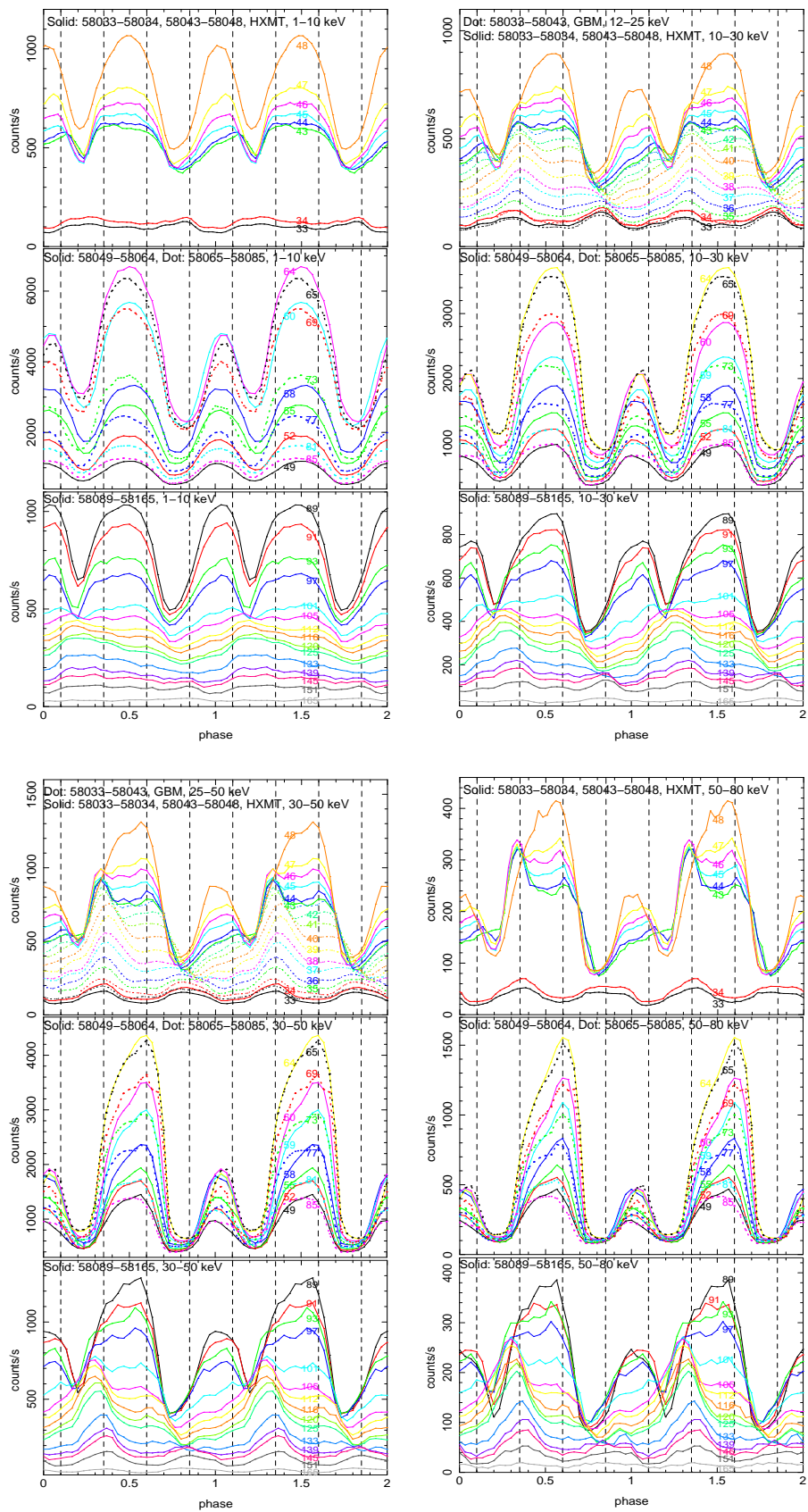

\begin{center}
   \includegraphics[width=2.4in,height=1.55in]{ZZ_1_10keVA.ps}
   \vspace{-0.40cm}
   \includegraphics[width=2.4in,height=1.55in]{ZZ_10_30keVA.ps}
   \includegraphics[width=2.4in,height=1.55in]{ZZ_1_10keVC.ps}
   \includegraphics[width=2.4in,height=1.55in]{ZZ_10_30keVC.ps}
\end{center}
   \vspace{-0.77cm}
\begin{center}
   \hspace{-6.17cm}
   \includegraphics[width=2.45in,height=1.58in]{ZZ_1_10keVD.ps}
\end{center}
   \vspace{-4.48cm}
\begin{center}
   \hspace{6.22cm}
   \includegraphics[width=2.45in,height=1.58in]{ZZ_10_30keVD.ps}
\end{center}

\begin{center}
   \includegraphics[width=2.4in,height=1.55in]{ZZ_30_50keVA.ps}
   \vspace{-0.40cm}
   \includegraphics[width=2.4in,height=1.55in]{ZZ_50_80keVA.ps}
   \includegraphics[width=2.4in,height=1.55in]{ZZ_30_50keVC.ps}
   \includegraphics[width=2.4in,height=1.55in]{ZZ_50_80keVC.ps}
\end{center}
   \vspace{-0.77cm}
\begin{center}
   \hspace{-6.17cm}
   \includegraphics[width=2.45in,height=1.58in]{ZZ_30_50keVD.ps}
\end{center}
   \vspace{-4.51cm}
\begin{center}
   \hspace{6.22cm}
   \includegraphics[width=2.45in,height=1.58in]{ZZ_50_80keVD.ps}
\end{center}
	\caption{Time evolution of the pulse profile of Swift J0243 in 1-10 keV (top left),
10-30 keV (top right), 30-50 keV (bottom left) and
50-80 keV (bottom right) during its 2017 giant outburst. 
The observed date of the profile since MJD 58000 are marked near the corresponding 
profiles. Dotted lines of 12-25 keV 
and 25-50 keV within MJD 58033-58043 
are from GBM data while others are from HXMT data. 
The high double peaks are 0.25 phase offset from the low double peaks,
indicating a 90 degree change of emission pattern.
}
\end{figure*}

\begin{figure}
	\includegraphics[width=3.0in]{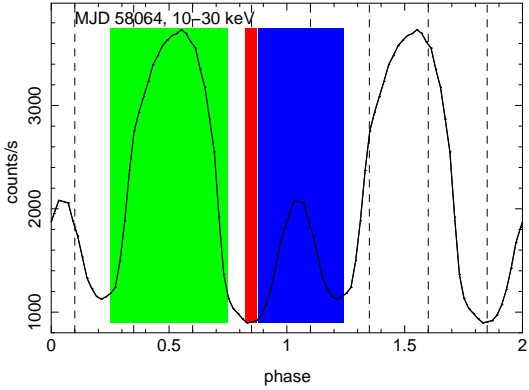}
	\vspace{-0.6cm}
	\caption{Illustration of the selected intervals of the main peak (green),
 the faint peak (blue), and the trough (red) for the 10-30 keV profile on MJD 58064.
	} 
\end{figure}

\begin{figure*}
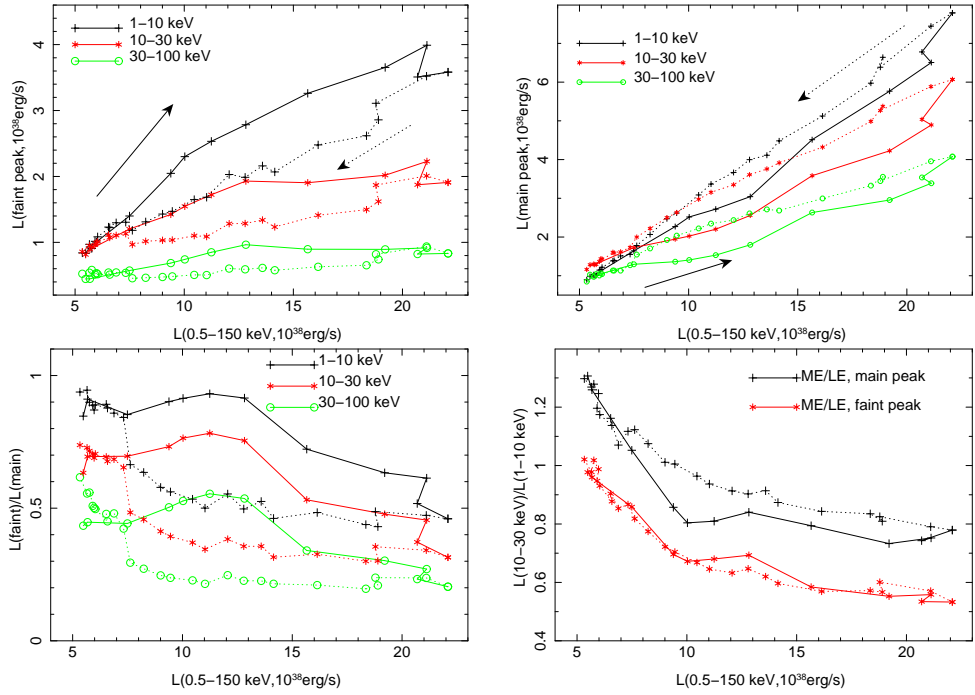

	\includegraphics[width=2.6in]{ZZmf3.ps}
	\vspace{-0.3cm}
	\includegraphics[width=2.6in]{ZZmf2.ps}
	\vspace{-0.3cm}
	\includegraphics[width=2.6in]{ZZmf4.ps}
	\includegraphics[width=2.6in]{ZZmfHR.ps}
	\caption{The pulsed luminosity of the faint peak (top left) and the main peak (top right)
	within different energy bands against the phase-averaged luminosity.
Bottom panels: the faint-to-main ratio and the hardness ratio.
The solid line indicates the rising stage, while the dotted line for the fading stage.
	} 
\end{figure*}

\section{Observation data}

{\it Insight}-HXMT is a Chinese X-ray satellite launched in 2017. There are 
three collimated instruments sensitive to low energy 
(LE, 1-15 keV), medium energy (ME, 5-30 keV), and high energy
(HE, 20-250 keV), with 
effective areas of 384, 952, and 5100 cm$^2$, respectively. 
For details of HXMT we refer to \citet{Zhang20} and references 
therein. HXMT monitored the giant outburst of Swift J0243 almost with a daily 
frequency except
for one week (MJD 58035-58042) at the beginning and provided a rich dataset 
for studies of super-Eddington accretion.

The Gamma-ray Burst Monitor \citep[GBM,][]{GBM09} on-board the \fm\  spacecraft 
is continuously monitoring the spin histories of X-ray
pulsars \citep{Fin09,Mal20}. 
For the date (MJD 58035-58042) without HXMT observations, 
we compiled the GBM pulsed profile data. 
Because the {\it Fermi}/GBM data is dominated by background and
the constant component of GBM light curves has been subtracted, 
only the pulsed profile is obtained.
The GBM pulsed profile is added with the \sw/BAT flux (multiplied 
by a constant), which is stretched by another constant. 
The two constants are adjusted 
to make the GBM profile matching that of HXMT on MJD 58043.
The selected GBM bands are 12-25 keV and 25-50 keV, roughly matching 
the 10-30 keV and 30-50 keV bands adopted for HXMT data.

\section{Temporal results}

To obtain phase-coherent pulse profiles, as did in \citet{Sug20}, we assign a pulse phase $\phi(t)$
to the event time $t$ as 
$\phi(t)=\int_{t_0}^{t}\nu(t)dt$,
where $\nu(t)$ is the pulse frequency interpolated from GBM measurements using 
a cubic spline function. The event arrival time $t$ is barycentric corrected using HXMT tool 
{\it hxbary} and is binary corrected using {\it BinaryCor} routine in Remeis 
ISISscripts\footnote{http://www.sternwarte.uni-erlangen.de/isis} with 
the orbital parameters obtained by the GBM pulsar
team\footnote{https://gammaray.msfc.nasa.gov/gbm/science/pulsars}. 

\subsection{Evolution of the pulse profile}

The pulse profiles of Swift J0243 in 1-10 keV, 10-30 keV, 30-50 keV, and 
50-80 keV bands are presented in Figure 1.
We first discuss the evolution of pulse profile in 10-30 keV.
At the beginning of the outburst (MJD 58033-58034, the black and red lines 
in the first panel), the profile was dominated by two peaks.
As the fluxes
increased with time, the peak around 0.35 increased faster than that around 0.85. 
Around MJD 58037, the peak around 0.85 disappeared, 
a flat plateau appeared around phase 0.7, and the whole profile 
looked like single-peaked. 
This is the double-to-single transition 
around $1.5\times10^{38}$erg\,s$^{-1}$ reported in previous studies.

As the fluxes increased further, both the main peak around 0.35 and the newly emerged plateau 
increased together. 
Around MJD 58043-58044, the original peak around 0.35 was mixed with the plateau, and
the mixed shape looked like a broad asymmetric bump. 
Meanwhile, a new, minor peak, appeared around phase 0.15.
On MJD 58048 (the orange line),
the signature of the original peak around 0.35 was disappeared, the peak shape
was more regular, and the faint peak around phase 1 was also more symmetric. 
This is the single-to-double transition
around $5\times10^{38}$erg\,s$^{-1}$ reported in previous studies.
Note that on MJD 58048, the two peaks were located at the troughs of the initial profile
on MJD 58033-59034.
That is, the maximum of the emission pattern changed about 90 degree, compared with 
that at the beginning of the outburst.

When the fluxes took off to the peak of the outburst within MJD 58049-58064
(the right second panel),
no apparent change of the peak location was observed anymore.
The two peaks around 0.5 and 1.0 increased with different rates. 
They have similar heights on MJD 58049, but the ratio of the maximum flux 
of the main peak to that of the faint peak is close to 2 on MJD 58064, 
the peak date of the outburst.

During the fading stage of the outburst, the trend described above reversed.
Around MJD 58091-58093, the shape of the main peak 
became bump-like again, and the faint peak became less symmetric.
Around MJD 58105 (the pink line in the right third panel), the peak feature around 
phase 0.3 reappeared and became the peak of the profile,
the previous faint peak around 1.0 disappeared,
and the whole profile looked single-peaked again. 
After that, both the two features around 
0.25 and 0.6 declined with decreasing fluxes together.
Note that during this single-peak regime, the location of the reappeared 
main peak is a little earlier than that during the rising state.
It is not certain whether this offset 
is real or due to some uncertainty in the phase alignment, but 
it will not affect the main results. 
As the fluxes decreased further more,
a second peak around phase 0.8 reappeared around MJD 58139, and the whole profile was 
double-peaked again.

Therefore, one can identify two low double-peak periods 
and one high double-peak period (MJD 58048-58090).
The high double peaks are 0.25 phase offset from the low double peaks.
In between is the single-peak regime, which changed to/from the high double-peak regime 
around $5\times10^{38}$erg\,s$^{-1}$ during both the rising and fading stage.
The single-peaked profile is a mix of the 
outburst-peak (high) feature (which is relatively lower at this time) 
and the initial low profile, for
which only the initial main peak around phase 0.35 increased with fluxes.
During this single-peak regime, the main peak of the high feature around phase 0.6
is more dominated over the faint one. The combination of features around phase 
0.35 and 0.6 (which is relatively lower) made the profiles looked single-peaked.

The pulse profiles of 1-10 keV generally follow the similar evolution trend as those of
10-30 keV, but some remarkable differences can be observed.
The phase position of the maximum flux of 1-10 keV profile
is generally a little earlier than that of 10-30 keV profile. For example,
the centroid of the 1-10 keV main peak on MJD 58064 
is around phase 0.5, while that of 10-30 keV profile is
around phase 0.55. During the high double-peak period, the 1-10 keV profiles 
look more symmetric than those of 10-30 keV profiles, and the 1-10 keV 
continuum level of the profile changed with a larger factor than 
that of 10-30 keV.

The pulse profiles of 30-50 keV follow a similar evolution trend as those of 
10-30 keV.
During MJD 58043-58047, the main peak feature of 30-50 keV around 
phase 0.35 is more prominent than that of 10-30 keV.
During the high double-peak regime, 
the constant level of the pulse profile of 30-50 keV was almost unchanged. 
The main peak around phase 0.6 is more dominated over the faint peak 
compared to that in 10-30 keV. The peak position of the 
30-50 keV profiles are around phase 0.6, which is about 0.05 phase later than 
that of the 10-30 keV profile.
Meanwhile, the shape of the main peak of 30-50 keV is 
right-titled compared with that in 10-30 keV band.

The pulse profiles of 50-80 keV are very similar with those of 30-50 keV.
During MJD 58043-58047, the main peak feature of 50-80 keV around 
phase 0.35 is also more prominent than that of 30-50 keV.
That is, the higher the energy, the more prominent the 0.35 peak.
Such a behavior is also true during the decline state.
During the high double-peak regime, the centroids of the main peak of 50-80 keV
are similar with those of 30-50 keV, but the shapes look more right titled 
than those of 30-50 keV.
The ratio of the maximum flux 
of the main peak to that of the faint peak of 50-80 keV is larger than 3 on MJD 58064. 
We also investigated the profiles of 80-100 keV and found they are similar to
those in 50-80 keV.

\subsection{Saturation of the faint peak in 10-100 keV}

During the high double-peak regime,
we can identify a transition of the behavior of pulse profile in 30-50 keV 
and 50-80 keV bands around MJD 58058.
As indicated by the blue line in the fifth panel of Figure 1,
the faint peak around phase 1 already reached the maximum level on MJD 58058. 
After MJD 58058, the main peak around phase 0.6 continued to increase until the 
outburst peak on MJD 58064, but the faint peak got saturated and stopped to increase 
any more. 

Such a transition is better illustrated with the luminosity of the main and faint 
peak. The low-energy profile of Swift J0243 showed a significant 
contribution from non-pulsed emission, part of which is due to reprocessed emission as 
indicated by the Fe \Ka emission line \citep{Tao19,Jai19}.
To reduce the effect of reprocessed and un-pulsed emission,
we calculate the pulsed-only luminosity of the main and faint peak, respectively.
The selected phase interval for 
the main peak is about 0.5, while that for the faint peak is
about 0.4. As an illustration, the selected intervals on MJD 58064 
are plotted in Figure 2 for 10-30 keV band. 
The spectra of each phase interval (also for the trough region) are extracted
and fitted with a model composed of an absorbed black body and
a cutoff power-law, by taking those around the
profile trough as background. 
The pulsed luminosities of the faint and main peak in three different bands
during the high double-peak regime (MJD 58048-58090) 
are plotted in the top two panels of Figure 3. Since the pulse profiles 
in 30-50 keV, 50-80 keV, and 80-100 keV are similar, we have combined 
these three bands into one of 30-100 keV.
The unabsorbed 0.5-150 keV luminosity including un-pulsed emission is adopted 
as the horizontal axis. To estimate the phase-averaged luminosity, 
a Gaussian model of Fe \Ka line is further added.
We adopt a fitting energy range of 1-150 keV and extrapolate 
the unabsorbed model to 0.5 keV. 

At the beginning of the high double-peak regime, 
the pulsed luminosity of the faint peak increased with the phase-averaged luminosity,
and the pulsed luminosity in 1-10 keV band had a faster increasing rate
than that in higher energy bands. 
The increasing rate in 1-10 keV band slowed down above $1\times10^{39}$erg\,s$^{-1}$,
and the faint peak got saturated in 10-30 keV and 30-100 keV bands above 
$1.3\times10^{39}$erg\,s$^{-1}$, as shown by the red and green lines in the top 
left panel of Figure 3.. 
During the fading stage, the pulsed luminosities of the faint peak were generally
smaller than those of the rising stage
and they are similar to those of the rising stage only below $7.5\times10^{38}$erg\,s$^{-1}$.
The pulsed luminosity of the main peak also had a faster increasing rate in 1-10 keV band than 
in other bands below
$1\times10^{39}$erg\,s$^{-1}$, but the main peak kept a similar increasing rate in all
three bands above that.

The ratio of the pulsed luminosity of the faint peak to the main peak is plotted 
in the bottom left panel of Figure 3. The faint-to-main ratio 
changed hardly below $1.3\times10^{39}$erg\,s$^{-1}$,
but above that the faint-to-main ratio decreased with increasing luminosity.
During the fading stage, the faint-to-main ratios were generally smaller than those of
the rising stage, and they jumped back to the level 
of the rising stage below $7.5\times10^{38}$erg\,s$^{-1}$.
The hardness ratio of the pulsed luminosity in 10-30 keV and 1-10 keV bands (ME/LE)
for the main and faint peak are plotted in the bottom right panel. 
The hardness ratio became softer with increasing luminosity 
for both the main and faint peak
below $1\times10^{39}$erg\,s$^{-1}$, but above that, it changed much less.

\section{Flattening of $\dot{\nu}$-$L$ relation}

In the standard scenario of disk accretion, the spin-up rate ($\dot{\nu}$) 
increases with luminosity ($L$). If the geometry of accretion flow changes somehow, 
a change of $\dot{\nu}-L$ relation will be expected. 
We calculate $\dot{\nu}$ by interpolating the GBM-measured spin-up rates 
on the corresponding HXMT observation time of Swift J0243. 
The calculated spin-up rate
against the phase-averaged luminosity is plotted in Figure 4. 
A flattening of $\dot{\nu}-L$ relation above $1.3\times10^{39}$erg\,s$^{-1}$ during the 
rising stage is clearly observed.
The $\dot{\nu}$ within the luminosity range 
of $5-13\times10^{38}$erg\,s$^{-1}$ during the rising stage can be fitted with a power-law 
model of $\dot{\nu}=1.15(\pm0.07)\times10^{-11} L_{38}^{1.02\pm0.03}$ Hz\,s$^{-1}$,
which is plotted as the blue dash line in Figure 4.
In contrast, the $\dot{\nu}$ within $1.3-2.2\times10^{39}$erg\,s$^{-1}$
follows a relation of $\dot{\nu}\propto L_{38}^{0.66\pm0.04}$.
During the fading stage, $\dot{\nu}$ are generally a little
smaller than those of the rising stage above $7\times10^{38}$erg\,s$^{-1}$.
We note that previous studies of $\dot{\nu}-L$ relation \citep[e.g.][]{Dor18,Zhang19}
did not distinguish the rising and fading stage and thus failed to reveal the
linearity of $\dot{\nu}$-L relation below $L_t$ and the flattening above $L_t$.

\begin{figure}
	\hspace{-0.2in}
	\includegraphics[width=3.2in]{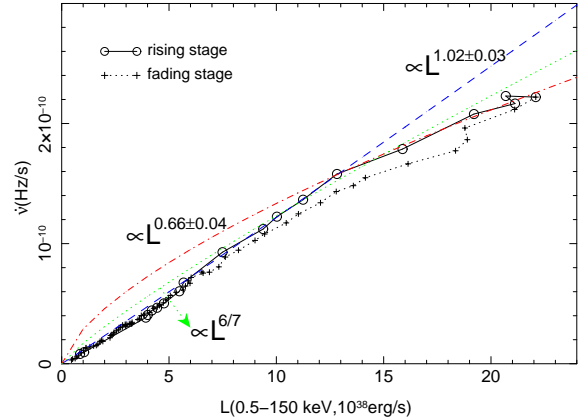}
	\caption{
Spin-up rate vs luminosity for Swift J0243. The data points within
$5-13\times10^{38}$erg\,s$^{-1}$ during the rising stage can be fitted
with a power-law of $\dot{\nu}\propto L^{1.02\pm0.03}$ (the blue dash line),
while those within $1.3-2.2\times10^{39}$erg\,s$^{-1}$ are fitted
with a power-law of $\dot{\nu}\propto L^{0.66\pm0.04}$ (the red dot-dash line).
The classical relation of $\dot{\nu}\propto L^{6/7}$
is also plotted (the green dot line). 
The linear relation below $1.3\times10^{39}$erg\,s$^{-1}$ is constant with a RPD disk.
The statistical errors of both $\dot{\nu}$ and $L$ are generally better than 0.1\%,
but their estimation is affected by the short effective exposure ($\sim2$ ks)
for each HXMT observation. We adopt 1\% uncertainty of $\dot{\nu}$ for the fitting,
and the quoted errors are for 90\% confidence level.
}
\end{figure}

\begin{figure*}
	   \includegraphics[angle=-90,width=2.8in,trim=100 130 100 130,clip]{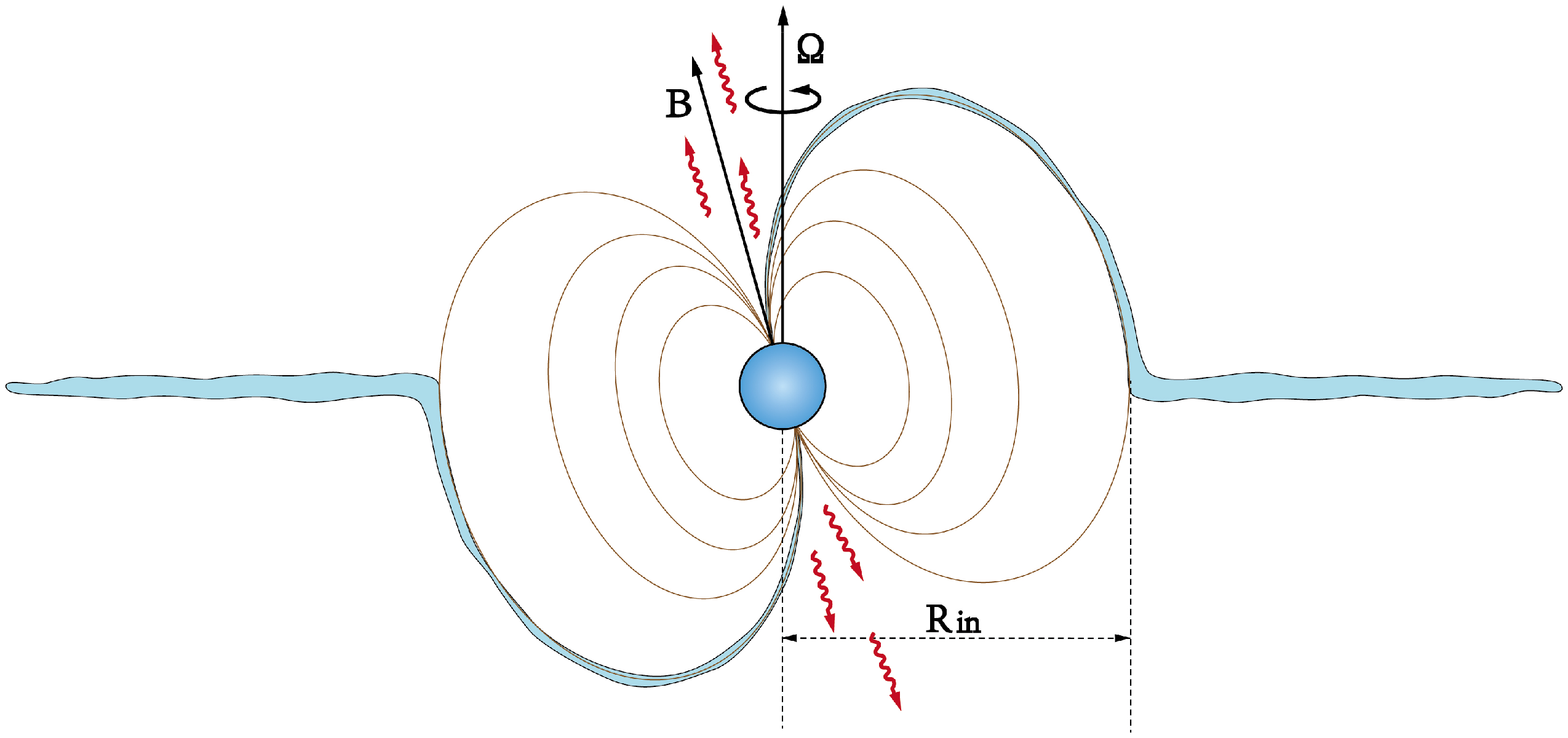}
			   \includegraphics[angle=-90,width=2.8in,trim=100 130 100 130,clip]{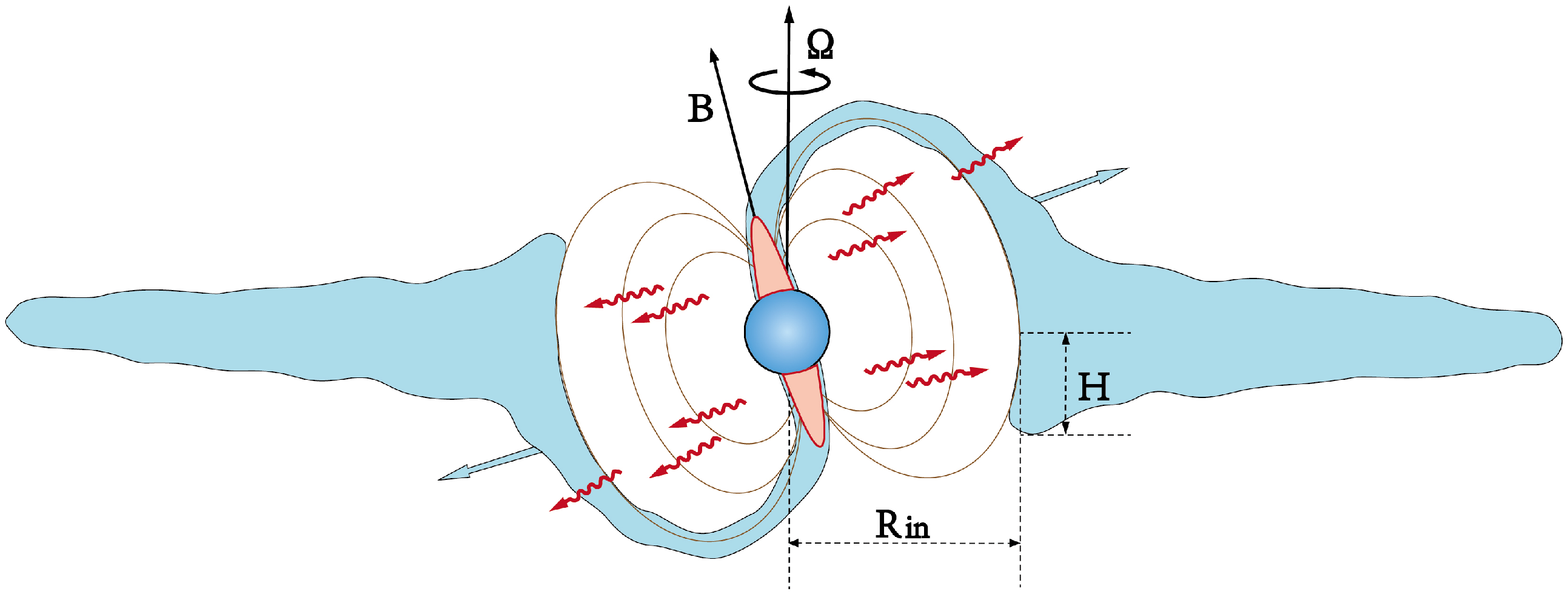}
				\caption{Illustration of the pencil beam along the magnetic pole
				at low state with a thin disk (left) and the fan beam perpendicular to the magnetic pole
				(accretion column) at high state with a thick disk (right),
				for which the irradiation of disk by the central accretion
				column above a certain luminosity ($L_t$) could change the accretion disk geometry 
				and cause a wind loss.}
\end{figure*}

\section{Discussion and conclusion}

We performed a detailed, phase-coherent pulse profile evolution study 
of Swift J0243 with HXMT and GBM data.
We found that the high double-peak profile is 0.25 phase offset from the 
low double-peak profile, 
and the single-peak-looked profile in between is a mix of the low double-peak profile 
and the infant profile of the high double-peak feature.
During the rising stage of the high double-peak regime, the faint peak 
increased with the phase-averaged luminosity below a luminosity of
$L_t\sim1.3\times10^{39}$erg\,s$^{-1}$,
but above $L_t$, the faint peak got saturated in 10-100 keV band.
The hardness ratio (ME/LE) of both the main 
and faint peak became softer with increasing luminosity below $L_t$, but showed 
much less changes with luminosity above $L_t$.
During the rising stage of the high double-peak regime, 
the $\dot{\nu}-L$ relation follows a linear correlation below $L_t$, and flattens above $L_t$.

In the standard accretion scenario of magnetized neutron star \citep[e.g.][]{GS73,Dav73,BS76}, 
at low accretion rates, the free-falling flow is stopped by nucleon collisions
at the surface of the neutron star, and X-ray radiation is emitted in a
pencil beam along the magnetic pole; above a critical luminosity,
the accretion flow was decelerated through a radiation shock, below which
the flow slowly settles down and forms an accretion column,
and X-ray photons are emitted mainly from the sidewall of the column in a fan
beam (see Figure 5).
Therefore, the 0.25 phase offset between the high and low double-peak profiles
is a strong evidence for a transition from a low pencil beam to a high fan beam
and for the existence of accretion column during the high double-peak regime.
The feature of the fan beam of accretion column first appears around a luminosity of
$1.5\times10^{38}$erg\,s$^{-1}$ and then totally dominates the profile above 
$5\times10^{38}$erg\,s$^{-1}$.
So we refer the high double-peak regime (MJD 58048-58090) as accretion column regime.

The saturation of the faint peak in 10-30 keV and 30-100 keV above $L_t$ is unexpected.
For the pencil beam at low fluxes, the main peak corresponds to the location
where the magnetic pole has the smallest inclination angle to the line of sight;
while for the fan beam from accretion column, the two peaks correspond to
the locations 90 degree before and after the smallest inclination angle.
The different faint-to-main ratios and phase lags between different energy bands
indicate that the emission in 1-10 keV and 10-100 keV bands could have different patterns.
The faster increasing rate of both the main and faint peak in 1-10 keV band
below $L_t$ also indicates that the 1-10 keV emission could arise from
a column region different from the 10-100 keV emission, which should mainly from
Comptonized emission near the base of the column\citep{LS88,Pou13}.
The saturation of the faint peak in 10-100 keV above $L_t$ could be related to
a change of the base of accretion column where the gravitational bending effect should play 
a role. The exact mechanism of the saturation of the faint peak is not clear 
to us currently. 

It is especially interesting to note that the saturation of the faint peak 
in 10-100 keV above $L_t$ is coincident with a flattening of the $\dot{\nu}-L$
relation above $L_t$, as shown in Figure 4. Since the $\dot{\nu}-L$ 
relation reflects the transfer of angular momentum to the neutron star, which is 
related to the geometry of accretion flow, a flattening of $\dot{\nu}-L$ 
relation above $L_t$ indicates that the saturation of the faint peak
might be related with a change of accretion flow.

The fitted power-law index around 1 of the $\dot{\nu}-L$ relation 
within $5-13\times10^{38}$erg\,s$^{-1}$ is apparently 
different from the value of 6/7 predicted by the standard disk accretion
model \citep{RJ77,GL79}, but is consistent with the model prediction of 
a radiation pressure dominated (RPD) disk \citep{Cha17,Cha19}.
Their calculation of a RPD disk showed that an increasing accretion rate
leads to an increase in disk thickness and the pressure balance 
can be satisfied at a same radius for different accretion rate. 
As a result, the magnetosphere size of a RPD disk
is almost independent on the mass accretion rate \citep[$\dot{m}$,][]{Cha17}:
\begin{equation}
	R_{in}\simeq3.6\times10^7\mu_{30}^{4/9}cm
\end{equation}
where $\mu_{30}$ is the magnetic moment in units of $10^{30}$G\,cm$^3$.
Then, $\dot{\nu}\propto\dot{m}$, following
$2\pi I\dot{\nu}=\dot{m}\sqrt{GMR_{in}}$, where $I$ and $M$ are the moment of inertia
and mass of the neutron star, and $G$ the gravitational constant.
Such a linear dependence is consistent with the observed linear $\dot{\nu}-L$ relation
below $L_t$ if the observed luminosity $L\propto\dot{m}$.

\citet{Dor20} has interpreted the single-to-double profile transition of Swift J0243
around $5\times10^{38}$erg\,s$^{-1}$ as an indication of
a possible transition of a gas-state disk to a RPD state.
As mentioned before, the single-to-double profile transition marks the time
when the profile was totally dominated by a fan beam and an accretion column fully 
came into being, and it does not necessarily mean a transition to a RPD state.
Nevertheless, the linear $\dot{\nu}-L$ relation within $5-13\times10^{38}$erg\,s$^{-1}$ 
we found here is consistent with the existence of a RPD disk at these luminosities.

From the observed $\dot{\nu}$ around $L_{t}$ one can infer
an inner radius of disk $R_{in}\sim1.2\times10^8$\,cm, assuming
the observed luminosity $L=\eta\dot{m}c^2$ with $\eta=0.2$ and $I=1.1\times10^{45}$g\,cm$^2$.
The thickness of a RPD disk around $L_t$ is $H\sim1\times10^7$cm \citep{SS73}, which could be
enhanced by a factor of $\sqrt{5}$ for the disk vertical structure considered
by \citet{Cha17,Cha19}.
A thickness ratio of $H/R_{in}\sim0.2$ does
correspond to the RPD regime as modeled in \citet{Cha17}.
From Eq. 1, one can estimate a magnetic field of $B\sim1.5\times10^{13}$\,G for a neutron star
of 1.4 $M_\odot$ with a radius of 10 km. 
Such a magnetic field is consistent with the observed critical luminosity for the
formation of accretion column and the maximum luminosity of
magnetized neutron stars estimated by \citet{Mus15a,Mus15b}.

The flattening of the observed $\dot{\nu}-L$ relation indicates
a less efficient transfer of angular momentum, which may be due to 
a loss of angular momentum and/or
a change of the geometry of the accretion flow above $L_t$.
If the faint peak was not saturated above $L_t$, even larger luminosity 
will be expected above $L_t$, and the flattening trend will be enhanced.
One possible scenario of the flattening of $\dot{\nu}-L$ relation is through
the irradiation of the disk by the central accretion column, which inflates the RPD disk,
and some angular momentum is lost in a wind \citep{Cha19}.
If this is true, the transition luminosity of $L_t$ represents a turning-on point of
significant radiation feedback of the column. Such a scenario is consistent with the 
possible ultrafast outflow
in Swift J0243 revealed by a {\it Chandra} observation on MJD 58068 \citep{Eij19},
when the luminosity ($\sim1.9\times10^{39}$erg\,s$^{-1}$) is above $L_t$.
In principle, the significant radiation feedback above $L_t$ could cause a change of accretion flow
and may lead to the observed saturation of the faint peak. 
Detailed modeling is needed to test whether the observed flattening of 
$\dot{\nu}-L$ relation and the saturation of the faint peak is really related
and to reveal how the faint peak get saturated above $L_t$.

In summary, the behavior of the pulse profile from an accretion column is complex.
The profiles show different shapes and phase-lags for different energy bands, and 
one peak could get saturated above a certain luminosity, which may be related with  
significant feedback of the column on the accretion flow, as indicated by a simultaneous 
flattening of $\dot{\nu}-L$ relation. These results provide
a basic ground for future modeling of the formation and evolution of accretion column
and for the study of super-Eddington accretion of ULXP.

\section*{Acknowledgements}
We thank our referee for helpful comments.
This work made use of data from the {\it Insight}-HXMT mission, a project
funded by China National Space Administration
(CNSA) and Chinese Academy of Sciences (CAS), and data from \fm/GBM and \sw/BAT.
This work is supported by National Key R\&D Program of China (2021YFA0718500), 
Natural Science Foundation of China (U1938113, 
12173103, U1838201, U1838202, U1938101 and 11733009),
and the Scholar Program of Beijing Academy of Science and Technology (DZ BS202002). 

\section*{Data Availability}
The data underlying this article are publicly available at http://archive.hxmt.cn.

\bibliographystyle{mn2e}

\end{document}